\documentclass[twocolumn,prl,aps]{revtex4}

\begin{document}

\title{\bf Environment Induced Entanglement in Markovian Dissipative Dynamics}

%\author{Fabio Benatti$^{a,b}$, Roberto Floreanini$^{b}$, 
%Marco Piani$^{a,b}$\\
%\small${}^a$Dipartimento di Fisica Teorica, Universit\`a di Trieste,
%Strada Costiera 11,\\
%\small 34014 Trieste, Italy\\
%\small ${}^b$Istituto Nazionale di Fisica Nucleare, Sezione di Trieste,
%34100 Trieste, Italy} 

\author{Fabio Benatti$^{a,b}$, Roberto Floreanini$^{b}$, 
Marco Piani$^{a,b}$}
\affiliation{${}^a$Dipartimento di Fisica Teorica, Universit\`a di Trieste, 
Strada Costiera 11, 34014 Trieste, Italy}
\affiliation{${}^b$Istituto Nazionale di Fisica Nucleare, Sezione di Trieste}

%\date{\null}

%\begin{document}

%\maketitle

%\vskip 2cm

\begin{abstract}
\noindent
We show that two, non interacting $2$-level systems, immersed in a common bath,
can become mutually entangled when evolving according to a Markovian, completely positive 
reduced dynamics.
\end{abstract}

\maketitle

%\vfill\eject

The role of quantum entanglement is of primary importance in 
quantum information and computation theory.
In recent years, a lot of research has been devoted to study how to entangle 
two systems by means of a direct interaction between them 
(see for instance~\cite{za,ci,du,kr,zy}).
In such a context, the presence of an environment, {\it e.g.} a generic noisy
reservoir or a heat bath, is commonly thought as counteracting entanglement 
creation, because of its decohering and mixing-enhancing effects.

However, a heat bath can also provide an indirect interaction between 
otherwise totally decoupled subsystems and thus a means to entangle them.
Indeed, this has been explicitly shown in a simple, exactly solvable 
model~\cite{bra}.
There, correlations between two subsystems are established during a 
transient phase where the reduced dynamics of the subsystems contains memory 
effects.

Instead, in this paper, we study the possibility that entanglement be created 
by the bath during the Markovian regime through a purely noisy mechanism.
We consider two, non interacting $2$-level systems,
weakly coupled to a common heat bath. 
We then start with a total Hamiltonian of the form
\begin{equation}
\label{1}
H_{\rm{tot}}=H_0^{(1)}+H_0^{(2)}+H_B+H_{\rm{int}}\ ,
\end{equation}
where $H_0^{(1)}$, $H_0^{(2)}$ and $H_B$ drive the dynamics of the two subsystems
and the bath in absence of each other; the interaction term couples
each subsystem independently with the bath, and can be taken of the form:
\begin{equation}
\label{2}
H_{\rm{int}}=\sum_{\alpha=1}^3\, (\sigma_\alpha\otimes{\bf 1})
\otimes V_\alpha\
+\ \sum_{\alpha=4}^6\, ({\bf 1}\otimes\sigma_{\alpha-3})\otimes V_\alpha\ ,
\end{equation}
where $\sigma_1$, $\sigma_2$, $\sigma_3$ are the Pauli matrices.
Notice that
we allow the subsystems to interact with the bath through different operators 
$V_\alpha$, while any direct coupling among themselves has been excluded.

In the weak-coupling limit~\cite{dav,go,lind,sp,al,bp}, the
reduced dynamics of the two $2$-level systems takes on a Markovian form.
Assuming a factorized initial state $\rho\otimes\rho_B$, where $\rho$ is a
state of the two subsystems and $\rho_B$ is an equilibrium state of
the bath, $\rho$ evolves in time according to
a quantum dynamical semigroup of completely positive maps with 
generator of the Kossakowski-Lindblad form:
\begin{equation}
\label{3}
\partial_t\rho(t)= -i\big[H\, ,\rho(t)\big]\, +\, L[\rho(t)]\ .
\end{equation}
The unitary term is the commutator with an effective Hamiltonian,
$H=H^{(1)}+H^{(2)} + H^{(12)}$,
consisting of single system pieces, including bath induced Lamb shifts,
\begin{equation}
\label{4}
H^{(1)}=\sum_{i=1}^3 H_i^{(1)} (\sigma_i\otimes {\bf 1})\ ,\quad
H^{(2)}=\sum_{i=1}^3 H_i^{(2)} ({\bf 1}\otimes \sigma_i)\ ,
\end{equation}
plus, possibly, a bath generated two-system coupling term
\begin{equation}
\label{5}
H^{(12)}=\sum_{i,j=1}^3 H_{ij}^{(12)} (\sigma_i\otimes\sigma_j)\ .
\end{equation} 
The dissipative contribution $L[\rho(t)]$ is as follows,
\begin{equation}
\label{6}
L[\rho]=\sum_{\alpha,\beta=1}^6\,
{\cal D}_{\alpha\beta}\ \Bigg[ 
{\cal F}_\alpha\, \rho\ {\cal F}_\beta\, -\, {1\over 2}
\Bigl\{{\cal F}_\beta {\cal F}_\alpha\,,\,\rho\Bigr\}\Bigg]\ ,
\end{equation}
with ${\cal F}_\alpha=\sigma_\alpha\otimes{\bf 1}$ for $\alpha=1,2,3$,
${\cal F}_\alpha={\bf 1}\otimes \sigma_{\alpha-3}$ for $\alpha=4,5,6$,
and ${\cal D}={\cal D}^\dagger$ a positive $6\times 6$ matrix which guarantees the complete
positivity of the evolution.
By writing
\begin{equation}
{\cal D}=\pmatrix{
  A & B \cr
  B^{\dagger} & C}\ ,
\label{7}
\end{equation} 
with $3\times 3$ matrices $A=A^\dagger$, $C=C^\dagger$ and $B$,
$L[\rho]$ assumes a form more amenable to a physical interpretation:
\begin{eqnarray}
\nonumber
&&\hskip -.5cmL[\rho]\!=\!\!
\sum_{i,j=1}^3\!\Bigg(\!
A_{ij}\Bigg[(
\sigma_i\otimes{\bf 1})\,\rho\,(\sigma_j\otimes{\bf 1})
-\frac{1}{2}\Big\{(\sigma_j\sigma_i\otimes{\bf 1})\,,\,\rho\Big\}
\Bigg]\\
\nonumber
&&\hskip .5cm
+C_{ij}\Bigg[(
{\bf 1}\otimes\sigma_i)\,\rho\,({\bf 1}\otimes\sigma_j)
           -\frac{1}{2}\Big\{({\bf 1}\otimes\sigma_j\sigma_i)\,,\,\rho\Big\}
\Bigg]\\
\label{8}
&&\hskip .5cm
+B_{ij}\Bigg[(\sigma_i\otimes{\bf 1})\,\rho\,({\bf 1}\otimes\sigma_j)
           -\frac{1}{2}\Big\{(\sigma_i\otimes\sigma_j)\,,\,\rho\Big\}
\Bigg]\\
&&\hskip .5cm
+B^*_{ij}\Bigg[({\bf 1}\otimes\sigma_j)\,\rho\,(\sigma_i\otimes{\bf 1})
           -\frac{1}{2}\Big\{(\sigma_i\otimes\sigma_j)\,,\,\rho\Big\}
\Bigg]
\Bigg)\ .
\nonumber
\end{eqnarray}
A generator of this form has been applied in quantum optics
to describe the phenomenon of collective resonance fluorescence
({\it e.g.} see \cite{pu}).
In the above expression, the first two contributions are dissipative 
terms that affect the first, respectively the second, system in absence 
of the other. On the contrary, the last two pieces represent the way 
in which the noise  may correlate the two subsystems; this effect
is present only if the matrix $B$ is different from zero.
\medskip

\noindent
{\bf Remark 1}\quad
From the rigorous derivation of Markovian master equations
\cite{dav,go}, 
one knows that the hamiltonian terms (\ref{4}), (\ref{5}) and
the entries of the matrix $\cal D$ in~(\ref{6}) contain
integrals of
two point time-correlation functions of bath operators: 
${\rm Tr}[\rho_B\,V_\alpha\,V_\beta(t)]$.
In particular, the matrices $[H_{ij}^{(12)}]$ in (\ref{5}) 
and $[B_{ij}]$ in (\ref{8}) do not vanish 
only if the bath state $\rho_B$ 
correlate bath-operators $V_\alpha$ pertaining to different subsystems, 
that is, if
the expectations ${\rm Tr}[\rho_B\,V_\alpha\,V_\beta(t)]$ are nonzero
when $1\leq\alpha\leq 3$ and
$4\leq\beta\leq 6$. 
Only in this case, entanglement has a chance to be created by the action of the bath.
Indeed, if $H^{(12)}=\,0$ and $B=\,0$, the two subsystems evolve independently 
and initially separable states may become more mixed, but certainly not entangled.
\medskip

\noindent
In order to check whether the reduced two-system density matrix $\rho$ 
gets entangled at time $t$ because of 
the time-evolution generated by equation~(\ref{3}), one can use the 
{\it partial transposition} criterion~\cite{pe,horo}:

\begin{quote}
{\sl if $\rho(t)$ acted upon with the partial
transposition with respect to one of the two subsystems 
has negative eigenvalues, then it is entangled;
in the $4$-dimensional case we are studying, also the reciprocal is true, 
namely if $\rho(t)$ 
is entangled, then, partial transposition makes negative 
eigenvalues appear.}
\end{quote}

\noindent
In physical terms, the bath is not able to create 
entanglement if and only if the partial transposition preserves 
the positivity of the state $\rho(t)$ for all times.
\medskip

\noindent
{\bf Remark 2}\quad
Strictly speaking, this criterion allows us 
to study the possibility of creating entanglement
starting from separable initial states.
When the initial state is already entangled, the partial transposition 
criterion cannot settle the question; in such cases, the analysis of the
entangling power of the bath can only be addressed through the study of how
entanglement measures evolve in time under dissipative reduced dynamics.
This problem requires a separate treatment and will not be addressed here.
\medskip

We therefore take separable states as initial states: 
as we shall see, this is not really a limitation
for the purpose of discussing the possibility of bath-induced entanglement creation.
Further, we can restrict our study to pure states; indeed, 
if the bath cannot create entanglement out of these, it will certainly not 
entangle their mixtures.
In view of this, we will consider initial states of the form
\begin{equation}
\label{9}
\rho(0)=\vert a_1\rangle\langle a_1\vert\otimes 
\vert b_1\rangle \langle b_1\vert\ ,
\end{equation}
where $\{\vert a_i\rangle\}$, $\{\vert b_i\rangle\}$, $i=1,2$, are orthonormal 
bases in
the two dimensional Hilbert spaces of the two subsystems.
For sake of definiteness, we will operate the partial transposition over the second
factor with respect to the basis $\{\vert b_1\rangle, \vert b_2\rangle\}$.

One can act with the partial transposition on both sides of 
equation (\ref{3}) and recast the result as
\begin{equation}
\label{10}
\partial_t\tilde{\rho}(t)=-i\Big[\widetilde{H}\,,\,\tilde{\rho}(t)\Big]\,+\,
\widetilde{L}[\tilde{\rho}(t)]\ ;
\end{equation}
here, $\tilde{\rho}(t)$ denotes the partially transposed matrix
$\rho(t)$, while $\widetilde{H}$ is a new Hamiltonian to which both the unitary and
the dissipative term in (\ref{3}) contribute
\begin{eqnarray}
\nonumber
\widetilde{H}=\sum_{i=1}^3 H_i^{(1)}&& {\hskip -.4cm} (\sigma_i\otimes {\bf 1})
+\sum_{ij=1}^3 H_i^{(2)} S_{ij} ({\bf 1}\otimes\sigma_j)\\
&&+\sum_{ij=1}^3 {\cal I}m\big(B\cdot S\big)_{ij} (\sigma_i\otimes\sigma_j)\ ,
\label{11}
\end{eqnarray}
where $S$ is the diagonal $3\times3$ matrix given by: $S={\rm diag}(-1,1,-1)$.
The additional piece $\widetilde{L}[\,\cdot\,]$ is of the form~(\ref{6}), but
with a new matrix ${\cal D}\to {\cal S}\cdot \widetilde{\cal D}\cdot {\cal S}$,
where
\begin{eqnarray}
&&\widetilde{\cal D}=
\pmatrix{A& {\cal R}e(B)+iH^{(12)}\cr
 {\cal R}e(B^T)-iH^{(12)}{}^T & C^T},\qquad
\label{12a}\\
&&{\cal S}=\pmatrix{{\bf 1}_3&0\cr0&S}\ ,
\label{12b}
\end{eqnarray}
and ${}^T$ denotes full transposition, while
$H^{(12)}$ is the coefficient matrix in~(\ref{5}). 
\medskip

\noindent
{\bf Remark 3}\quad
Although $\tilde{\rho}(t)$ evolves according to a master equation formally
of Kossakowski-Lindblad form, the new coefficient matrix $\widetilde{\cal D}$
need not be positive.
As a consequence, the time-evolution generated by (\ref{10}) may result to be neither
completely positive, nor positive and therefore may not preserve 
the positivity of the initial state $\tilde{\rho}(0)\equiv \rho(0)$.
\medskip

Notice that both the hamiltonian and the dissipative terms of the original
master equation (\ref{3}) contribute to the piece 
$\widetilde{L}[\,\cdot\,]$ in (\ref{10}), the
only one that can produce negative eigenvalues.
In particular, this makes more transparent the physical mechanism according to
which a direct hamiltonian coupling $H^{(12)}$ among the two systems can
induce entanglement: on $\tilde{\rho}(t)$, $H^{(12)}$ ``acts'' as a dissipative
contribution, which in general does not preserve positivity.
The entanglement power of purely hamiltonian couplings have been
extensively studied in the recent literature~\cite{za,ci,du,kr,zy}. 
Instead, in the following
we shall concentrate our attention on whether entanglement can be
produced by the purely dissipative action of the heat bath;
henceforth, we shall ignore the contribution of the matrix
$H^{(12)}$ in $\widetilde{\cal D}$. In other words, we shall
take into account only baths for which the induced two-system hamiltonian
coupling in (\ref{5}) is vanishingly small.%
\footnote{Physically speaking, this is not really a limiting choice.
As already observed, the coefficients $H^{(12)}_{ij}$ in (\ref{5})
depend on bath correlation functions 
$G_{\alpha\beta}(t)={\rm Tr}[\rho_B V_\alpha(t) V_\beta(0)]$
for which $\alpha=1,2,3$ and $\beta=4,5,6$. Then, one easily checks that in the
singular coupling limit derivation~\cite{go} of the master equation (\ref{3}), the
contribution $H^{(12)}$ vanishes for the physically relevant case
of time-symmetric bath correlations.
Instead, in the weak coupling limit~\cite{dav}, 
a real $G_{\alpha\beta}(t)$ would suffice
to assure the condition $H^{(12)}=\,0$.}
\medskip

\noindent
{\bf Remark 4}\quad
If $\widetilde{D}$ is positive, then, 
the time-evolution generated by (\ref{10})
is completely positive; therefore, $\tilde{\rho}(t)$ is positive at all times
and entanglement is not created.
Instances of baths for which this happens can easily be
provided:
\begin{enumerate}
\item
$B=\,0$: in such a case,  $\widetilde{\cal D}$ is positive since
such are $A$ and $C^T$, due to the positivity of $\cal D$;
this corresponds to a bath that
does not dynamically correlate the two
subsystems;
\item
${\cal R}e(B)=\,0$: as before, $\widetilde{\cal D}$ is block-diagonal and thus
positive;
\item
${\cal I}m(B)=\,0$ and $C^T=C$ or $A^T=A$: in the first case, $\widetilde{\cal D}={\cal D}$,
while in the second $\widetilde{\cal D}={\cal D}^T$;
\item
$A^T=A$ and $C^T=C$: in this case 
$\widetilde{\cal D}=({\cal D}+{\cal D}^T)/2$.
\end{enumerate}
\noindent
In the last three cases, despite the fact that the two subsystems are 
now dynamically correlated by the bath, the effect is not sufficient for 
entanglement production. Further, notice that entanglement is not
created also in baths for which the corresponding coefficient matrix $\cal D$ can be written
as a convex combination of matrices satisfying the previous conditions.
\medskip

In order to check the presence of negative eigenvalues in $\tilde{\rho}(t)$,
instead of examining the full equation (\ref{10}) we find convenient to study the quantity
\begin{equation}
{\cal E}(t)=\langle\psi\vert\, \tilde{\rho}(t)\, \vert\psi\rangle\ ,
\label{13}
\end{equation}
where $\psi$ is any $4$-dimensional vector.
Assume that an initial separable state $\tilde{\rho}$ has indeed developed a 
negative eigenvalue at time $t$, but not before.
Then, there exists a vector state $\vert\psi\rangle$ and a time $t^*<t$
such that ${\cal E}(t^*)=0$, ${\cal E}(t)>0$ for $t< t^*$ and
${\cal E}(t)<0$ for $t> t^*$. 
The sign of entanglement creation may thus be given by a negative
first derivative of ${\cal E}(t)$ at $t=t^*$.
Moreover, by assumption, the state $\rho(t^*)$ is separable.
Without loss of generality, one can 
set $t^*=0$ and, as already remarked, restrict 
the attention to factorized pure initial states.

In other words, the two subsystems, initially prepared in a 
state $\rho(0)=\tilde\rho(0)$ as in (\ref{9}), will become entangled
by the noisy dynamics induced by their independent interaction with the bath
if: 1) ${\cal E}(0)=\,0$ and 2) $\partial_t {\cal E}(0)<0$,
for a suitable vector $|\psi\rangle$,
\begin{equation}
|\psi\rangle=\sum_{i,j=1}^2 \psi_{ij}\, |a_i\rangle\otimes|b_j\rangle\ .
\label{14}
\end{equation}
Given (\ref{9}), condition 1) readily implies: $\psi_{11}=\, 0$.
\medskip

\noindent
{\bf Remark 5}\quad Note that entanglement creation can not be detected
by looking at the sign of the first derivative of ${\cal E}(t)$ unless
the test vector $|\psi\rangle$ is entangled itself. Indeed, 
${\cal E}(t)$ is never negative for a separable $|\psi\rangle$.
Thus, both components $\psi_{12}$ and $\psi_{21}$ in (\ref{14})
have to be different from zero, since otherwise $|\psi\rangle$ becomes separable.
\medskip

\noindent
{\bf Remark 6}\quad When $\partial_t{\cal E}(0)>0$ for all choices 
of the initial state $\rho(0)$ and probe vector $|\psi\rangle$, 
the bath is not able to entangle the two systems, since $\tilde\rho$
remains positive. The treatment of the case $\partial_t{\cal E}(0)=\,0$
requires special care: in order to check entanglement creation,
higher order derivatives of $\cal E$, possibly with a time dependent
$|\psi\rangle$, need to be examined.
\medskip

In order to prove that indeed there are baths for which ${\cal E}(0)=\,0$ and 
$\partial_t{\cal E}(0)$
is negative, let us first make the choice $|a_1\rangle=|b_1\rangle=|+\rangle$
and $|a_2\rangle=|b_2\rangle=|-\rangle$, where $|\pm\rangle$ are the
eigenstates of $\sigma_3$; the general case is considered below.
For $|\psi\rangle=(|+\rangle\otimes|-\rangle + |-\rangle\otimes|+\rangle)/\sqrt{2}$,
one finds
\begin{equation}
\partial_t {\cal E}(0)={\rm Tr}\big[ {\cal D}\, {\cal R}\big]\ ,
\label{15}
\end{equation}
where $\cal D$ is as in (\ref{7}), while
\begin{equation}
{\cal R}=\pmatrix{
  P & Q \cr
  Q & P}\ ,\quad
P={1\over2}\pmatrix{\phantom{-}1 & i & 0\cr
                    -i & 1 & 0\cr
                    \phantom{-}0 & 0 & 0\cr}\ ,
\label{16}
\end{equation}
and $Q={\rm diag}(-1/2,1/2,0)$. 
Although $P$ is a projector, $(2Q)^2={\rm diag}(1,1,0)$, 
and as a consequence ${\cal R}$ possesses
one negative eigenvalue, $(1-\sqrt{2})/2$, of multiplicity two.
Any bath for which the Kossakowski coefficient matrix ${\cal D}$
has support only in the negative eigenspace of $\cal R$ would 
generate a negative $\partial_t {\cal E}(0)$, and therefore
entangle the initially separated state
$\rho(0)=|+\rangle\langle +|\otimes |+\rangle\langle+|$.

A simple explicit example in which this happens is given by the following
two-parameter matrix $\cal D$, with
\begin{equation}
A=C=\pmatrix{1 & -ia & 0\cr
             ia & 1 & 0\cr
             0 & 0 & 0\cr},\ \,
B=\pmatrix{b & \phantom{-}0 & 0\cr
           0 & -b & 0\cr
           0 & \phantom{-}0 & 0\cr}\ ,
\label{17}
\end{equation}
where $a$ and $b$ are real constants.%
\footnote{This example can be easily generalized by adding more parameters;
in these cases however, the description of the region in parameter space
for which entanglement is generated becomes more involved.}
Positivity of $\cal D$, required by
the complete positivity of the subsystems Markovian dynamics (\ref{3}),
is guaranteed by $a^2+b^2\leq 1$. Inside this unit disk,
the region for which $\partial_t{\cal E}(0)$ in (\ref{15}) is negative 
is characterized by the condition $a+b>1$. Actually, by changing
the initial state $\rho(0)$ and the probe vector $|\psi\rangle$,
one can show that entanglement is created in all four disk portions outside
the embedded square $|a\pm b|\leq 1$. Notice that inside this square  
$\widetilde{\cal D}$ is positive, so that there the time evolution 
of the partially transposed density matrix $\tilde\rho(t)$ generated by
(\ref{10}) is also completely positive: in this case, entanglement can not be 
created for any choice of the initial state $\rho(0)$ and of the vector $|\psi\rangle$.

Now that we have shown that a Markovian dynamics can indeed entangle
the two subsystems via a purely noisy mechanism, let us discuss
in more detail the condition for entanglement creation.
Although in general the basis vectors $|a_i\rangle$, $|b_i\rangle$, 
introduced in (\ref{9}),
are not eigenstates of $\sigma_3$, they can always
be unitarily rotated to the basis $|\pm\rangle$:
\begin{eqnarray}
&&|a_1\rangle= U |+\rangle \qquad |a_2\rangle= U |-\rangle\ ,
\nonumber\\
&&|b_1\rangle= V |+\rangle \qquad |b_2\rangle= V |-\rangle\ .%\label{18}
\end{eqnarray}
The unitary transformations $U$ and $V$ induce orthogonal transformations
$\cal U$ and $\cal V$, respectively, on the Pauli matrices:
\begin{equation}
U^\dagger \sigma_i U=\sum_{j=1}^3 {\cal U}_{ij}\sigma_j\ ,\quad
V^\dagger \sigma_i V=\sum_{j=1}^3 {\cal V}_{ij}\sigma_j\ .
\label{19}
\end{equation}
With these definitions, for a generic separable initial state (\ref{9})
and arbitrary vector $|\psi\rangle$ such that ${\cal E}(0)=\,0$, the condition 
$\partial_t {\cal E}(0)<0$ for entanglement formation can be expressed
as the following expectation value over the product of $6\times6$ matrices:
\begin{equation}
\vec{w}^\dagger\cdot \Big[ \Psi^\dagger\, {\cal W}^T\, \widetilde{\cal D}\,
{\cal W}\, \Psi \Big]\cdot \vec{w}<0\ ,
\label{20}
\end{equation}
where $\widetilde{\cal D}$ is as in (\ref{12a}) (with $H^{(12)}$
set to zero as explained before), 
while the remaining matrices are given by:
\begin{equation}
{\cal W}=\pmatrix{ {\cal U} & 0\cr
                     0   &{\cal V} }\ ,\quad
\Psi=\pmatrix{ \psi_{21}\, {\bf 1}_3 & 0\cr
                0 & -\psi_{12}\, {\bf 1}_3}\ ,
\label{21}
\end{equation}
and the components of the 6-vector $\vec{w}$ by
the Pauli matrix elements:
\begin{equation}
w_i=\langle +| \sigma_i |-\rangle\ ,\quad w_{i+3}=w_i^*\ ,\quad i=1,2,3\ .
\label{22}
\end{equation}

A more manageable condition for checking entanglement production can be obtained 
by noticing that (\ref{20}) is quadratic in the components $\psi_{12}$ and $\psi_{21}$
of $|\psi\rangle$. By suitably rearranging the expression in (\ref{20}),
one can then show that entanglement is generated if 
the following inequality, independent from the probe vector $|\psi\rangle$,
holds:
\begin{equation}
\langle u | A | u \rangle \, \langle v | C^T | v \rangle <
\big|\langle u | {\cal R}e(B) | v \rangle \big|^2\ ;
\label{23}
\end{equation}
the 3-vectors $|u\rangle$ and $|v\rangle$ are not completely arbitrary:
they contain the information about the starting factorized state (\ref{9}),
and their components can be \hbox{expressed as}:
\begin{equation}
u_i=\sum_{j=1}^3 {\cal U}_{ij}\, w_j\ ,\quad
v_i=\sum_{j=1}^3 {\cal V}_{ij}\, w_j^*\ .
\label{24}
\end{equation}
Therefore, a given bath will be able to entangle the two subsystems evolving
with the Markovian dynamics generated by (\ref{3}) and characterized by
the Kossakowski matrix (\ref{7}), if there exists an initial
state $|a_1\rangle\langle a_1|\otimes |b_1\rangle\langle b_1|$,
or equivalently orthogonal transformations $\cal U$ and $\cal V$,
for which the inequality (\ref{23}) is satisfied.

The condition (\ref{23}) can thus be used to check the entangling power
of specific Markovian time evolutions. As an example, consider a bath
leading to a Kossakowski matrix (\ref{7}) for which $A=B=C$; this
choice corresponds to a special case of collective resonance fluorescence
\cite{aga,pu}.
Provided the hermitian matrix $A$ is not symmetric, one can easily prove 
that there are initial states of the form (\ref{9}) with $|a_1\rangle=|b_1\rangle$
that will get entangled by the noisy dynamics. Indeed, in this case condition (\ref{23})
reduces to:
\begin{equation}
\big|\langle u|\, {\cal I}m(A) \, |u\rangle\big|^2>0\ ,
\label{25}
\end{equation}
which is clearly satisfied for any $|u\rangle$ outside the null eigenspace
of ${\cal I}m(A)$. When $A$ is real however, (\ref{25}) is violated and entanglement is
not created, since the partial transpose state $\widetilde{\rho}(t)$
evolves in time with a completely positive dynamics.

\centerline{------------------}

After completion of the manuscript, our attention has been drawn
to Refs.\cite{ki,sch,ba,ja} which have connections with the topics discussed
in this letter.

\end{document}